\documentclass[12pt]{article}
\usepackage{amsmath,amsthm,amscd,amsfonts,amssymb}
\usepackage{latexsym}
\newcommand{\CC}{\mathbb C}

\newcommand{\hh}{\mathcal H}
\newcommand{\kk}{\mathcal K}

\newcommand{\mm}{{\bf a}}
\newcommand{\bb}{{\bf b}}
\newcommand{\pp}{{\bf P}}
\newcommand{\qq}{{\bf Q}}
\newcommand{\xx}{{\bf X}}
\newcommand{\XX}{{\bf x}}
\newcommand{\xsmall}{{\bf x}}
\newcommand{\yy}{{\bf Y}}
\newtheorem{thm}{Theorem}[section]

\newtheorem{cor}[thm]{Corollary}

\begin{document}
\title{An Entropic Uncertainty Principle for Quantum Measurements} 
\author{{\it M Krishna  and K R Parthasarathy} \\ 
Institute of Mathematical Sciences\\ Taramani,
Chennai 600 113, India \\
{\small e-mail:  krishna@imsc.ernet.in  and   krp@imsc.ernet.in}
\\
{\it \small Dedicated to the memory of D. Basu}
}
\date{}
\maketitle
\begin{abstract}
The entropic uncertainty principle as outlined by Maassen and Uffink in
\cite{MR89f:81027} for a pair of non-degenerate observables in a finite level quantum
system is generalized here to the case of a pair of arbitrary quantum 
measurements.  In particular, our result includes not only the case of projective 
measurements (or equivalently, observables) exhibiting degeneracy but also an 
uncertainty principle for a single measurement.
\end{abstract}

\section{Introduction}
In the context of quantum computation and information, the notion of a measurement 
for a finite level quantum system has acquired 
great importance.  (See, for example, Nielsen and Chuang \cite{MR1796805}).
Suppose that a finite level quantum system is described by pure states which are unit 
vectors in a $d$-dimensional complex Hilbert Space $\hh$ with scalar product
$<\cdot, \cdot>$ which is linear in the second variable.  By a measurement $\xx$ we 
mean $\xx = (X_1, X_2, \dots, X_m)$, a finite sequence of positive operators 
satisfying the relation $\sum_{i=1}^n X_i = I$.  If $\psi \in \hh$ is a unit 
vector, then (in the Dirac notation) $p_i = <\psi|X_i|\psi>, i=1, \dots, m$ is
a probability distribution on the set $\{1,2,\dots, m\}$ which is interpreted as a labeling of the
possible elementary outcomes of the measurement.  The corresponding uncertainty
involved in such a measurement is measured by the entropy
\begin{equation}
\label{eqn1}
H(\xx, \psi) = - \sum_{i=1}^m p_i ~ {\text log}_2 ~ p_i.
\end{equation}
Now consider two different measurements, $\xx = (X_1, X_2, \dots, X_m)$ and
$\yy = (Y_1, Y_2, \dots, Y_m)$ in the state $\psi$. 
We would then like to describe the entropic uncertainty principle by a sharp 
lower bound for the sum $H(\xx, \psi) + H(\yy, \psi)$ of the two entropies.
Such an approach for observables was first initiated by Bialynicki-Birula and Mycielski \cite{MR52:7403}.
Pursuing a conjecture of Kraus \cite{MR88e:81007}, Maassen and Uffink \cite{MR89f:81027}
obtained a sharp lower bound for the sum of entropies of two measurements $\xx$ and $\yy$ when all the $X_i$ and $Y_j$
are one dimensional projections, i.e., when $\xx$ and $\yy$ reduce to observables
without degeneracy.  Following the arguments of Maassen and Uffink \cite{MR89f:81027}
closely in using the Riesz-Thorin interpolation theorem and combining it with
an application of Naimark's theorem  \cite{Helstrom} as outlined in \cite{MR1810813}
we obtain a lower bound in the case of a pair of arbitrary measurements of a 
finite level system.  Our lower bound does coincide with the Maassen-Uffink
lower bound in the case of observables without degeneracy.

\vspace{.5cm}

{\noindent \bf Acknowledgement}  Part of this work was done at the Delhi center
of the Indian Statistical Institute when the second author was supported by the
Indian National Science Academy in the form of C. V. Raman Research Professorship.
The second author thanks both the I.S.I and INSA for their support.

\section{The Main Result}

We say that a measurement $\xx = (X_1, X_2, \dots, X_m)$ is {\it projective}
if each $X_i$ is an orthogonal projection.  In such a case one has
\begin{equation}
\label{eqn2}
X_iX_j = \delta_{ij} X_j ~~ {\text for all } ~~ i, j \in \{1, 2, \dots, m\}.
\end{equation}

\begin{thm}
\label{thm1}
Let $\pp = (P_1, P_2, \dots, P_m)$, $\qq = (Q_1, Q_1, \dots, Q_n)$ be two 
projective measurements and let $\psi$ be a pure state in $\hh$.  Then
\begin{equation}
\label{eqn3}
H(\pp,\psi) + H(\qq,\psi) \geq -2 ~{\text log}_2 ~~ \substack{{\text max}\\i,j}
\frac{|<\psi|P_iQ_j|\psi>|}{\|P_i\psi\| \|Q_j\psi\|} , 
\end{equation}
where, on the right hand side, the maximum is taken over all the $1\leq i\leq m$,
$1\leq j \leq n$ satisfying the conditions $P_i\psi \neq 0$, $Q_j\psi \neq 0$.

\end{thm}

Before proceeding to the proof of this theorem we shall present the well-known
Riesz-Thorin interpolation theorem in a convenient form. Let $T=((t_{ij})), ~
1\leq i \leq m, 1\leq j \leq n$ be any matrix of order $m\times n$ with entries
from the field $\CC$ of complex scalars.  In any space $\CC^k$ we define the 
norms
\begin{equation}
\label{eqn4}
\|\XX\|_p = 
\begin{cases}
& \left( \sum_{i=1}^k |x_i|^p\right)^{1/p} \hfill ~~ if ~~ 1\leq p < \infty,\\
&   \\
& \substack{{\text max} \\ 1 \leq i \leq k} ~~~~ |x_i|  \hfill ~ if ~~ p = \infty, 
\end{cases}
\end{equation}
where $\XX^{t} = (x_1, x_2, \dots, x_k)$.

Consider the operator $T:\CC^n \rightarrow \CC^m$ defined by 
\begin{equation}
\label{eqn5}
(T\xsmall)_i = \sum_{j=1}^n t_{ij} x_j
\end{equation}
and define
\begin{equation}
\label{eqn6}
\|T\|_{p,q} = \substack{\sup \\ \xsmall : \|x\|_p =1} ~~~ \|T\xsmall\|_q ~~ {\text where} ~~ \frac{1}{p} + \frac{1}{q} = 1.
\end{equation}

With these notations we have the following theorem.

\begin{thm}
\label{thm2}
Suppose $p_0, q_0, p_1, q_1$ are in the interval $[1, \infty]$ and 
$\frac{1}{p_0} + \frac{1}{q_0} = \frac{1}{p_1} + \frac{1}{q_1}=1$ and
\begin{equation}
\label{eqn7}
\|T\|_{p_0,q_0} \leq m_0 ,  ~~~ \|T\|_{p_1,q_1} \leq m_1. 
\end{equation}
Define $p_t, q_t$ for $0 < t < 1$ by
\begin{equation}
\label{eqn8}
\frac{1}{p_t} = t \frac{1}{p_1} + (1-t) \frac{1}{p_0},~~~ 
\frac{1}{q_t} = t \frac{1}{q_1} + (1-t) \frac{1}{q_0}.
\end{equation}
Then
\begin{equation}
\label{eqn9}
\|T\|_{p_t,q_t} \leq m_t ,  ~~ {\text where} ~~ m_t = m_0^{1-t}m_1^t,  
\end{equation}
for every $0< t < 1$.
\end{thm}

{\noindent \bf Proof: } This is a very special case of Theorem IX.17, 
pages 27-28 of Reed and Simon \cite{MR58:12429b}. \hfill \qed

{\noindent \bf Proof of Theorem \ref{thm1} :} Without loss of generality
we can assume that $P_i\psi \neq 0$, $Q_j\psi \neq 0$ for every
$1\leq i\leq m, 1\leq j \leq n$.  Otherwise, we can restrict the following
argument to the subset of indices which obey this condition.  Define

\begin{equation}
\label{eqn10}
\phi_i = \frac{P_i\psi}{\|P_i\psi\|}, ~~ 
\psi_i = \frac{Q_i\psi}{\|Q_i\psi\|} 
\end{equation}
and observe that $\{\phi_i\}$ and $\{\psi_j\}$ are orthonormal sets.  Put

\begin{equation}
\label{eqn11}
t_{ij} = <\phi_i | \psi_j>, ~~~ 1\leq i\leq m, 1\leq j \leq n.
\end{equation}
For any $\xsmall \in \CC^n$ we have, 
\begin{equation}
\label{eqn111}
\begin{split}
\sum_{i=1}^m | \sum_{j=1}^n t_{ij}x_j |^2 
&= \sum_{i=1}^m | <\phi_i | \sum_{j=1}^n x_j |\psi_j> |^2 \\ 
&\leq \| \sum_{j=1}^n x_j |\psi_j> \|^2  =  
 \sum_{j=1}^n |x_j |^2.   
\end{split}
\end{equation}
Thus the operator $T:\CC^n \rightarrow \CC^m$ defined by the matrix T
satisfies the inequality 
\begin{equation}
\label{eqn12}
\|T\|_{2,2} \leq 1.
\end{equation}

On the other hand
\begin{equation}
\label{eqn13}
\substack{{\text max}\\i} | \sum_{j=1}^n t_{ij}x_j | \leq 
\substack{{\text max}\\i,j} |t_{ij}| \sum_{j=1}^n |x_j|. 
\end{equation}
In other words, 
\begin{equation}
\label{eqn14}
\|T\|_{1,\infty} \leq R,  ~~ {\text where} ~~ R = \substack{{\text max}\\ i,j}
|t_{ij}|. 
\end{equation}
Now apply Theorem \ref{thm2} after putting
$$
p_0=q_0=2, p_1=1, q_1=\infty, m_0=1, m_1=R.
$$
Then we have, 
\begin{equation}
\label{eqn15}
\|T\|_{p_t,q_t} \leq R^t, ~~ 0 < t < 1,  
\end{equation}
where a  computation shows that $p_t = 2/(1+t)$ and $q_t = 2/(1-t)$.  Define
the vectors $\mm \in \CC^n, \bb \in \CC^m$ by
\begin{equation}
\label{eqn16}
a_j = <\psi_j|\psi>, j=1,2,\dots, n ~~~ 
b_i = <\phi_i|\psi>, i=1,2,\dots, m.
\end{equation}
We have 
\begin{equation}
\label{eqn17}
\begin{split}
(T\mm)_i &= \sum_{j=1}^n t_{ij}a_j \\
&= \sum_{j=1}^n <\phi_i|\psi_j><\psi_j|\psi> \\ 
&= \sum_{j=1}^n \frac{<\phi_i|Q_j\psi><Q_j\psi|\psi>}{\|Q_j\psi\|^2} \\ 
&= \sum_{j=1}^n <\phi_i|Q_j\psi> \\ 
&= <\phi_i | \sum_{j=1}^n Q_j |\psi> \\ 
&= <\phi_i | \psi> = b_i. \\ 
\end{split}
\end{equation}

By inequality (\ref{eqn15}) we now conclude that 

\begin{equation}
\label{eqn18}
\left(\sum_{i=1}^m |<\phi_i|\psi>|^{\frac{2}{1-t}}\right)^{\frac{1-t}{2}}
\leq 
R^t \left(\sum_{j=1}^n |<\psi_j|\psi>|^{\frac{2}{1+t}}\right)^{\frac{1+t}{2}},
\end{equation}
for every $0 < t < 1$. Denoting
$$
p_i = <\psi | P_i | \psi> = |<\phi_i|\psi>|^2, ~~ 
q_j = <\psi | Q_j | \psi> = |<\psi_j|\psi>|^2,  
$$
we see that the inequality (\ref{eqn18}) can be expressed as, after
raising both sides to power $2/t$ and trasfering the second factor on the
right hand side to the left,  
\begin{equation}
\label{eqn19}
\left(\sum_{i=1}^m p_i p_i^{\frac{t}{1-t}}\right)^{\frac{1-t}{t}}
\left(\sum_{j=1}^n q_j q_j^{-\frac{t}{1+t}}\right)^{-\frac{1+t}{t}}
\leq R^2,  ~~ 0 < t < 1.
\end{equation}
Taking natural logarithms, letting $t\rightarrow 0$ and using L'Hospital's rule
we get
$$
\sum_{i=1}^m p_i~~  {\text log}~~  p_i 
+ \sum_{j=1}^m q_j ~~ {\text log}~~  q_j  \leq 2 {\text log}~  R. 
$$
This completes the proof of the theorem. \hfill \qed

\begin{cor}
\label{cor1}
Let $\pp$ and $\qq$ be projective measurements and let $\psi$ be any pure state.  
Then
\begin{equation}
\label{eqn20}
H(\pp, \psi) + H(\qq, \psi) \geq - 2 ~~ {\text log} ~~ \substack{{\text max} \\ i, j} \|P_iQ_j\|.
\end{equation}
\end{cor}

{\noindent \bf Proof:} This is immediate from Theorem \ref{thm1} when we
note that 
\begin{equation}
\label{eqn21}
\begin{split}
|<\psi |P_iQ_j |\psi>| & = 
|<P_i \psi |P_iQ_j | Q_j \psi>| \\ 
& \leq \|P_iQ_j\| ~~ \|P_i\psi\| ~~ \|Q_j\psi\|.\hfill \qed 
\end{split}
\end{equation}
 
{\noindent \bf Remark:}  Inequality (\ref{eqn20}) becomes trivial, in the sense that the right hand
side vanishes,  if and only if $\|P_iQ_j\| =1$ for some $i, j$.  This, in turn, is equivalent to
finding a nonzero vector in the intersection of the ranges of $P_i$ and $Q_j$
for some $i,j$. 

One can also consider a mixed state of the form
$$
\rho = \sum_{i=1}^r \pi_i |\psi_i><\psi_i|,  ~~ \pi_i > 0, ~~ \sum_{i=1}^r \pi_i = 1,
$$
where $\psi_i, ~ i=1,2,\dots, r$ are unit vectors.  Then for any measurement
$\xx = (X_1,X_2, \dots, X_m)$ one obtains a probability distribution
$$
p_k = Tr(\rho X_k) = \sum_{i=1}^r \pi_i <\psi_i|X_k|\psi_i>, ~~ 1\leq k \leq m.
$$
We write 
$$
H(\xx, \rho) = - \sum_{k=1}^m p_k ~ {\text log}_2 ~ p_k.
$$
Then we note that $(p_1, p_2, \dots, p_m)$ is a convex combination of the 
probability distributions $(p_{i1}, p_{i2}, \dots, p_{im}), ~ 1\leq i \leq r$,
where
$$
p_{ik} = <\psi_i|X_k|\psi_i>, ~~ 1\leq k \leq m.
$$
If now $\pp$ and $\qq$ are two projective measurements it follows from the concavity 
property of entropy (see section 11.3.5, pages 516-518 of Nielsen and Chuang 
\cite{MR1796805}) that 
\begin{equation}
\label{eqn22}
\begin{split}
H(\pp,\rho) + H(\qq, \rho) &\geq \sum_{i=1}^r \pi_i 
\left[H(\pp, \psi_i) + H(\qq, \psi_i)\right] \\
&\geq -2 ~ {\text log} ~ \substack{{\text max}\\i, j} \|P_iQ_j\|.
\end{split}
\end{equation}
The importance of this inequality lies in the fact that the right hand side
is independent of the state $\rho$.

\begin{thm}
\label{thm3}
Suppose $\pp = (P_1, P_2, \dots, P_m)$ is a projective measurement and
$\yy = (Y_1, Y_2, \dots, Y_n)$ is an arbitrary measurement.  Then for
any pure state $\psi$, 
\begin{equation}
\label{eqn23}
H(\pp,\psi) + H(\yy, \psi) \geq  
 -2 ~ {\text log} ~ \substack{{\text max}\\i, j} 
\frac{|<\psi|P_iY_j|\psi>|}{\|P_i\psi\| ~~  \|Y_j^{\frac{1}{2}}\psi\|}.
\end{equation}
where the maximum is over all $i, j$ for which 
$P_i\psi \neq 0, Y_j^{1/2}\psi \neq 0$.
\end{thm}

{\noindent \bf Proof: }  We look upon $\yy$ as a positive operator valued 
measure on the finite set $\{1, 2, \dots, n\}$.  In an orthonormal basis
of $\hh$, the operators $P_i, Y_j, ~~ 1\leq i\leq m, ~1\leq j \leq n$ can all be 
viewed as positive semidefinite matrices.  By Naimark's theorem \cite{Helstrom} 
as interpreted in \cite{MR1810813} for finite dimensional Hilbert spaces
we can construct matrices of the form
\begin{equation}
\label{eqn24}
\tilde{Q_j} = \left[ \begin{matrix} Y_j & L_j \cr L_j^{\dagger} & Z_j \end{matrix}\right], 
~~ 1\leq j \leq n 
\end{equation}
so that $\tilde{Q_j}$ 's are projections in an enlarged Hilbert space
$\hh\oplus\kk$ where $\kk$ is also a finite dimensional Hilbert space and 
$$
\sum_{j=1}^n \tilde{Q_j} = I_{{}_{\hh\oplus\kk}}
$$
Define
\begin{equation}
\label{eqn25}
\begin{split}
\tilde{P_1} &= \left[ \begin{matrix} P_1 & 0 \cr 0 & I_{{}_\kk} \end{matrix}\right], 
\\ 
\tilde{P_i} &= \left[ \begin{matrix} P_i & 0 \cr 0 & 0 \end{matrix}\right], 
~~ 2\leq i \leq m \\
\tilde{\psi} &= \left[ \begin{matrix} \psi  \cr 0 \end{matrix}\right], 
\end{split}
\end{equation}
where the vectors in $\hh\oplus\kk$ are expressed as column vectors 
$[\substack{u\\v}]$ with $u \in \hh$ and $v \in \kk$. Then $\tilde{\psi}$ is a pure
state and $\tilde{\pp} = (\tilde{P_1}, \tilde{P_2}, \dots, \tilde{P_m})$, 
$\tilde{\qq} = (\tilde{Q_1}, \tilde{Q_2}, \dots, \tilde{Q_n})$ are
projective measurements in an enlarged system.  By Theorem \ref{thm1} we have
\begin{equation}
\label{eqn26}
H(\tilde{\pp},\tilde{\psi}) + H(\tilde{\qq}, \tilde{\psi}) \geq  
 -2 ~ {\text log} ~ \substack{{\text max}\\i, j} 
\frac{|<\tilde{\psi}|\tilde{P_i}\tilde{Q_j}|\tilde{\psi}>|}{\|\tilde{P_i}\tilde{\psi}\| ~~  \|\tilde{Q_j}\tilde{\psi}\|}.
\end{equation}
On the other hand we have
\begin{equation}
\label{eqn27}
\tilde{P_i}\tilde{\psi} = \left[ \begin{matrix} P_i\psi \cr  0 \end{matrix}\right], 
 ~~ \tilde{Q_j}\tilde{\psi} = \left[ \begin{matrix} Y_j\psi \cr L_j^{\dagger}\psi \end{matrix}\right]. 
\end{equation}
This implies
$$
<\tilde{\psi}|\tilde{P_i}\tilde{Q_j}|\tilde{\psi}> = <\psi|P_i~Y_j|\psi> ~~~
{\text and} ~~~
<\tilde{\psi}|\tilde{P_i}|\tilde{\psi}> = 
<\psi|P_i|\psi>.
$$
Since $\tilde{Q_j}$ is a projection we have
$$
\|\tilde{Q_j}\tilde{\psi}\|^2 = <\tilde{\psi}|\tilde{Q_j}|\tilde{\psi}> = 
<\psi|Y_j|\psi> = \|Y_j^{\frac{1}{2}}\psi\|^2.
$$
Thus (using the above two equations) inequality (\ref{eqn26}) reduces to inequality  (\ref{eqn23}).

\begin{thm}
\label{thm4}
Let $\xx = (X_1, X_2, \dots, X_m)$, 
$\yy = (Y_1, Y_2, \dots, Y_n)$ be two arbitrary measurements.  Then for
any pure state $\psi$, 
\begin{equation}
\label{eqn28}
H(\xx,\psi) + H(\yy, \psi) \geq  
 -2 ~ {\text log}_2 ~ \substack{{\text max}\\i, j} 
\frac{|<\psi|X_iY_j|\psi>|}{\|X_i^{\frac{1}{2}}\psi\| ~~  \|Y_j^{\frac{1}{2}}\psi\|}.
\end{equation}
where the maximum is over all $i, j$ for which 
$X_i^{1/2}\psi \neq 0, Y_j^{1/2}\psi \neq 0$.
\end{thm}

{\noindent \bf Proof: } As in the proof of Theorem \ref{thm3}, use Naimark's
theorem \cite{Helstrom} and construct the projections $\tilde{Q_j}$ as in equation (\ref{eqn24}).
Define
\begin{equation}
\label{eqn29}
\begin{split}
\tilde{X_1} &= \left[ \begin{matrix} X_1 & 0 \cr 0 & I_{{}_\kk} \end{matrix}\right], 
\\ 
\tilde{X_i} &= \left[ \begin{matrix} X_i & 0 \cr 0 & 0 \end{matrix}\right], 
~~ 2\leq i \leq m, \\
\end{split}
\end{equation}
and consider the state $\tilde{\psi}$ as defined by equation (\ref{eqn25}).
Then $\tilde{\qq} = (\tilde{Q_1}, \tilde{Q_2}, \dots, \tilde{Q_n})$ is a
projective measurement and $\tilde{\xx} = (\tilde{X_1}, \tilde{X_2}, \dots,
\tilde{X_m})$ is a measurement.  Hence by Theorem \ref{thm3}, 
\begin{equation}
\label{eqn30}
H(\tilde{\qq},\tilde{\psi}) + H(\tilde{\xx}, \tilde{\psi}) \geq  
 -2 ~ {\text log}_2 ~ \substack{{\text max}\\i, j} 
\frac{|<\tilde{\psi}|\tilde{Q_j}\tilde{X_i}|\tilde{\psi}>|}
{\|\tilde{X_i}^{\frac{1}{2}}\tilde{\psi}\| ~~  \|\tilde{Q_j}\tilde{\psi}\|}.
\end{equation}
As in the proof of Theorem \ref{thm3} we note that
$$
<\tilde{\psi}|\tilde{Q_j}|\tilde{\psi}> = \|\tilde{Q_j}\tilde{\psi}\|^2 = 
<\psi|Y_j|\psi> = \|Y_j^{\frac{1}{2}}\psi\|^2.
$$
Clearly, inequality (\ref{eqn30}) reduces to 
\begin{equation}
\label{eqn31}
H(\xx,\psi) + H(\yy, \psi) \geq  
 -2 ~ {\text log}_2 ~ \substack{{\text max}\\i, j} 
\frac{|<\psi|Y_jX_i|\psi>|}
{\|{Y_j}^{\frac{1}{2}}\psi\| ~~  \|{X_i}^{\frac{1}{2}}\psi\|},
\end{equation}
which is the same as equation (\ref{eqn28}) owing to the self-adjointness of
$X_i$ and $Y_j$. \hfill \qed
 
\begin{cor}
\label{cor2}
Let $\xx = (X_1, X_2, \dots, X_m)$, $\yy = (Y_1, Y_2, \dots, Y_n)$
be arbitrary measurements and let $\rho$ be any state.  then
\begin{equation}
\label{eqn32}
H(\xx,\rho) + H(\yy, \rho) \geq  
 -2 ~ {\text log}_2 ~ \substack{{\text max}\\i, j} 
\|{X_i}^{\frac{1}{2}}{Y_j}^{\frac{1}{2}}\|.
\end{equation}
\end{cor}

{\noindent \bf Proof: }  Owing to the concavity of Shannon entropy it is 
enough to prove the Corollary when $\rho$ is a pure sate determined by
a unit vector $\psi$.  Now the required result is immediate from the theorem
above if we observe that
\begin{equation}
\label{eqn33}
\begin{split}
|<\psi|X_iY_j|\psi>| & = |<X_i^{\frac{1}{2}}\psi | X_i^{\frac{1}{2}}Y_j^{\frac{1}{2}} | Y_j^{\frac{1}{2}}\psi>|\\
& \leq \|X_i^{\frac{1}{2}}Y_j^{\frac{1}{2}}\| ~~ \|X_i^{\frac{1}{2}}\psi\| ~~ \|Y_j^{\frac{1}{2}}\psi\|. \hfill \qed 
\end{split}
\end{equation}

{\noindent \bf Remark:}  Putting $\xx = \yy$ in inequality (\ref{eqn32}) we get
$$
H(\xx,\rho)  \geq  
 - ~ {\text log}_2 ~ \substack{{\text max}\\i, j} 
\|{X_i}^{\frac{1}{2}}{X_j}^{\frac{1}{2}}\|.
$$
This yields a nontrivial uncertainty principle even for a single measurement 
since the right hand side need not vanish.

\vspace{1cm}

{\noindent \bf Example:} Let $G$ be a finite group of cardinality $N$ and let
$\widehat{G}$ denote its dual space consisting of all the inequivalent
irreducible unitary representations of G.  Denote by $L^2(G)$, the
$N$-dimensional complex Hilbert space of all functions on $G$ with the
scalar product
$$
<f|g> = \sum_{x \in G} \overline{f(x)}g(x), ~~~~ f, g \in L^2(G).
$$
For any $\pi \in \widehat{G}$, let $d(\pi)$ denote the dimension of the 
representation space of $\pi$ and let $\{\pi_{ij}(\cdot), ~ 1 \leq i,j \leq d(\pi)\}$
denote the matrix elements of $\pi$ in some orthonormal basis of its
representation space.  From the Peter-Weyl theory of representations
we have two canonical orthonormal bases for $L^2(G)$:
\begin{enumerate}
\item $\{|x> = 1_{{}_{\{x\}}}, ~~ x \in G\};$
\item $\{\sqrt{\frac{d(\pi)}{N}}\pi_{ij}(\cdot),  ~~ 1\leq i,j\leq d(\pi), \pi \in \widehat{G}\}$,
\end{enumerate}
where $1_{{}_{\{x\}}}$ denotes the indicator function of the singleton set 
$\{x\}$ in $G$.  Consider the projective measurements 
$$
\qq = \{Q_x, x\in G\}, ~~ Q_x = |x><x|, ~~ \pp = \{P_{i,j,\pi}, ~ \pi \in 
\widehat{G}, 1\leq i,j\leq d(\pi)\}, 
$$
where 
$$
P_{i,j,\pi} = \frac{d(\pi)}{N} |\pi_{ij}><\pi_{ij}|.
$$
For any unit vector $\psi$ in $L^2(G)$, we have
\begin{equation}
\begin{split}
<\psi|Q_xP_{i,j,\pi}|\psi> & = \frac{d(\pi)}{N} <\psi|x><\pi_{ij}|\psi>\pi_{ij}(x), \\
\|Q_x\psi\|^2 &= <\psi|Q_x|\psi> = |\psi(x)|^2, \\
\|P_{i,j,\pi}\psi\|^2 &= \frac{d(\pi)}{N}|<\pi_{ij}|\psi>|^2.
\end{split}
\end{equation}
Thus our entropic uncertainty principle assumes the form 
\begin{equation}
\begin{split}
 & -\sum_{x\in G} |\psi(x)|^2 ~{\text log}_2 ~|\psi(x)|^2 - 
\sum_{\substack{1\leq i,j\leq d(\pi)\\ \pi\in \widehat{G}}} 
|\widehat{\psi}(i,j,\pi)|^2 ~{\text log}_2  ~~ |\widehat{\psi}(i,j,\pi)|^2 \\ 
& \geq -2 {\text log}_2  ~~ \substack{{\text max}\\i,j,\pi,x} \sqrt{\frac{d(\pi)}{N}}
 ~~ |\pi_{ij}(x)|,
\end{split}
\end{equation}
where
$$
\widehat{\psi}(i,j,\pi) = \sqrt{\frac{d(\pi)}{N}}<\pi_{ij}|\psi>
$$
is the (noncommutative ) Fourier transform of $\psi$ at the ${ij}^{th}$ entry of the
irreducible representation $\pi$.  since $\pi_{ij}(x)$ is the ${ij}^{th}$ 
entry of the unitary matrix $\pi(x)$ and $\pi(e) = I_{d(\pi)}$ at the 
identity element $e$ we have
$$
\substack{{\text max}\\i,j,\pi,x} |\pi_{ij}(x)| = 1. 
$$
Thus the entropic uncertainty principle reduces to 
\begin{equation}
\label{eqnlast}
\begin{split}
& -\sum_{x\in G} |\psi(x)|^2 ~ {\text log}_2 ~ |\psi(x)|^2 - 
\sum_{\substack{1\leq i,j\leq d(\pi)\\ \pi\in \widehat{G}}} 
|\widehat{\psi}(i,j,\pi)|^2 ~ {\text log}_2 ~ |\widehat{\psi}(i,j,\pi)|^2\\  
& \geq  {\text log}_2  ~~ N  -  {\text log}_2 ~~  
\substack{{\text max} \\ \pi \in \widehat{G}}  ~~ d(\pi), 
\end{split}
\end{equation}
for every unit vector $\psi \in L^2(G)$.  When $G$ is abelian every $\pi$ is one 
dimensional and the right hand side reduces to ${\text log}_2~  N$.  In this 
case, when $\psi(x) \equiv 1/\sqrt{N}$, the inequality in (\ref{eqnlast}) becomes
an equality.


\end{document}